Monte Carlo study of the ordering in a strongly frustrated liquid crystal

S. George[1,2], C. Bentham[1], X. Zeng[3], G. Ungar[3,4] and GA Gehring[1]

[1]Department of Physics and Astronomy, University of Sheffield, Sheffield S3 7RH, UK

[2] CERN, Route de Meyrin 385, 1217 Meyrin, Switzerland

[3] Department of Materials Science and Engineering, University of Sheffield,

Sheffield, S1 3JD

[4] Department of Physics, Zhejiang Sci-Tech University, 310018 Hangzhou, China

**Abstract**

We have performed Monte Carlo simulations to investigate the temperature dependence of the ordering of the side chains of the X-shape liquid crystal molecules which are arranged in a hexagonal array. Each hexagon contains six side chains, one from each side of the hexagon. Each liquid crystal molecule has two, dissimilar, side chains, one that contains silicon and one containing fluorine. Like chains attract each other more strongly than unlike chains and this drives an order-disorder transition. The system is frustrated because it is not possible to find a configuration in which all the hexagons are occupied by either all silicon or all fluorine chains. There are two phase transitions. If only pairwise interactions are included it is found that there is a novel fluctuating phase between the disordered phase and the fully ordered ground state. This did not agree with the experiments where an intermediate phase was seen that had long range order on one of the three sublattices. Agreement was found when the calculations were modified to include attractive three body interactions between the silicon chains.

**I Introduction**

The Monte Carlo technique is a very suitable tool for investigating novel phase transitions in classical systems because no assumptions need be made about the type of order expected and accurate predictions made of the order of the transition and, where appropriate, it may be used to evaluate the critical exponents [1,2]. It is particularly useful for systems that show frustration [3,4,5], which is an interesting and challenging topic, because it can lead to systems that have no long range order or partial order [6,7,8]. The model used to describe the system that we are reporting on here is highly unusual because a previous mean field analysis showed that in the presences of pair interactions only it had an intermediate phase that was partially ordered while having a fully ordered ground state [9].

An order –disorder transition occurs between two honeycomb-like hexagonal columnar liquid crystal phases of X-shaped molecules such as **A1** and **A2** (Figure 1a)**.** These molecules contain a rigid rod-like core with two incompatible lateral flexible chains. The rod-like terphenyl cores form the walls of the honeycomb, while the flexible side-chains fill the cells. For the compounds discussed in this paper, one of the lateral chains contains carbosilane groups (Si-chain) and the other is fluorinated hydrocarbon (F-chain) (Figure 1b).

Self-assembly of such compounds leads to multi-colour superlattices where the two kinds of lateral chains micro-phase separate into neighbouring columns with polygonal shaped cross sections [9,10,11]. Such columnar structures, when projected along the column axis, are two dimensional tilings of polygons. The borders of the *n*–sided polygons are formed from the rigid molecular cores end-linked by hydrogen bonds, and the interior of the tile is occupied by *n* chains. The number of side chains allowed within one tile is fixed by the volume constraints and the only degree of freedom is in the fraction of each type of chain in a given tile. This rich family of molecules exhibit a wide variety of structures, including a Kagome lattice with a mixture of hexagonal and triangular tiles, two coloured lattices with hexagonal or triangular tiles, multi-coloured tilings with up to five kinds of tiles of different shape and colour, and more recently a tiling of giant octagonal and square tiles [9,10,11].

We are particularly interested in the case where each X-molecule has two side chains where like side chains attract significantly more than unlike chains. This leads to an order-disorder transition where the fractions of the two types of chain occupy the tiles at random at high temperature but segregate below some transition temperature. The side chains together fill the volume allowed by the tiling but, as they are flexible, they intermingle. This leads us to a model where the science is dominated by the fraction of each type of chain within each tile and not by the particular arrangement of the chains among the sites around the tile.

The structure of the strongly frustrated liquid crystal is described in section II and the model is described in section III where we also draw attention to the similarities and differences to the triangular spin, $S =3$, model with antiferromagnetic interactions and to the spin ½ antiferromagnetic Kagome lattice. For pair interactions between all the chains within each hexagon we obtain a very important result, namely that there is only *one* relevant energy in the system which represents the difference between the attraction of each type of chain with others of the same type and the mutual repulsion between chains of a different type. This means that there is an exact symmetry between the two types of chains in all the allowed ordered states. As each liquid crystal molecule has one chain of each type the total number of chains of each type is equal and that furthermore any interchange of two side chains will change the occupation fractions in two neighbouring hexagons; this is analogous to Kawasaki dynamics for a magnetic problem.

The Monte Carlo methods are described in section IV and are used first to investigate a model where the chains interact only through pairwise interactions. The phase diagram that is obtained using only pairwise interactions between side chains indicates the existence of two phase transitions, a transition from the disordered state to a highly unusual strongly fluctuating phase, whose properties are investigated, and finally to an ordered phase.

We determine the effects of including three body interactions between the chains to the model in section V in order to break the symmetry between Si- and F-chains and show that this gives two definite transitions in agreement with earlier work [9]. Qualitative arguments suggest that the transition at the higher temperature might be in the same universality class as the 3-state Potts model. The critical exponent for the order parameter, $\beta$, obtained from the X-ray diffraction measurements, taken as ordering starts to occur, is compared to the approximate value obtained from simulation in section VI and that for the 3-state Potts model. Finally the conclusions are given in section VII.

**II The structure of the strongly frustrated liquid crystal**

The information about the liquid crystal including its basic molecular structure and the various ordering phases that will be discussed in this paper are shown in figure 1. The basic liquid crystal structure is formed from a backbone of terphenyl molecules with terminating glycerol end groups that self-organise into a hexagonal lattice that stack to form columns. Each terphenyl molecule has two side chains one of which contains carbosilane groups (Si-chain, coloured red), and the other is a semi perfluorinated hydrocarbon chain (F-chain, coloured blue), as shown in Fig. 1(a,b) forming an X-shaped molecule [9]. It may be observed that the points where the side chains are joined to the terphenyl molecules lie on a Kagome lattice.

Ordering is driven by the different interactions between the Si-chains and F-chains, in particular the attraction between like chains is significantly stronger than that between unlike chains. An order-disorder phase transition is observed as the temperatures is lowered from the disordered phase (one-colour) lattice, shown in Fig. 1(f) where the two kinds of lateral chains are mixed in each hexagon to one in which there is partial ordering, the two colour lattice, as shown in Fig. 1(e) where one sublattice is occupied predominantly by Si- chains and the other two sublattices are randomly mixed with a higher concentration of F-chains [9]. AFM as well as neutron scattering experiments have verified that the phase that we observed was as sketched in Fig 1(e), and that the equivalent phase with the ordering of the F- chains was not seen ([9] and in the Supplementary Information therein). The final transition to the fully ordered, three colour, phase Fig. 1(c, d) was not observed experimentally because the liquid crystal undergoes a transition to a three dimensional crystalline state before any second transition could be seen. As is clear from Fig. 1 it is not possible to have all the hexagons occupied either by all Si- or all F-chains. It is this frustration that gives rise to the very interesting phase diagram and critical properties. The lowest energy state, which is non-degenerate, has three sublattices two of which are fully occupied with one type of chain and the third is mixed equally. Note that in all these figures the number of Si- and F-chains summed over all cells is exactly equal.

Experimental results are given for two different compounds, **A1** and **A2** that have the same terphenyl molecule backbone, but the side chains are longer for **A1** than for **A2**.

The mean field theory, considered earlier [9], predicted that there should be two phase transitions. As the temperature is lowered it predicted there is first a transition from the disordered state to a partially ordered state as shown in Fig. 1(e), or equivalently one in which the F-chains were fully ordered on one sublattice, and then at a lower temperature a transition to the fully ordered state as shown in Figs. 1(c) and 1(d). The results of the Monte-Carlo simulations discussed in Section IV will be compared with this earlier prediction.

**III The model with pairwise interactions**

Within a given hexagon the six flexible side chains can mingle and hence a model is adopted where each side chain interacts equally with all the other side chains inside the same hexagon. This model neglects the fact that a given side chain might interact more with the chains from neighbouring chains than from chains fixed to the opposite sides of the hexagon. (Actually no transition would be observed at all if only nearest neighbour interactions were included because this would map on to an antiferromagnetic spin ½ Kagome lattice.)

The energy of the whole system is a sum over the energies of each hexagon. The energy of a given hexagon is given in terms of the pair interaction energies between all the chains within that hexagon in terms of the energy between two Si-chains, $S$, the energy between two F-chains, $F$, and the mixed interaction between one Si and one F-chain, $M$. This means there are *three* characteristic energies, $S$, $F$, and $M$. We note that we consider only a two dimensional model in which interactions between the ordering in one hexagon and the hexagons above and below it are neglected. The possibility that one hexagon contains 5 or 7 chains instead of 6 is discounted because this is sterically forbidden.

The energy of a given hexagon containing $n$ Si-chains and $(6-n)$ F-chains is found from the total number of distinct Si-chain−Si-chain interactions that exist, $\frac{n(n-1)}{2}$, and similarly for the F-chains, $\frac{(6-n)(5-n)}{2}$, and the number of interactions between Si-chains and F-chains, $n(6-n)$. The energies of a hexagon with $n$ Si-chains and $(6-n)$ F-chains and their multiplicities (the number of ways that $n$ Si-chains and $(6-n)$ F-chains can be arranged among the six sites) are given in the Table I; the multiplicities add to $64=2^6$ as expected. The Table also shows the largest statistical weight is attached to the configurations where the numbers of Si- and F-chains are almost equal so indicating the importance of the entropy in driving this transition.

We consider a flip of the chains between hexagons **a** and **b** as shown in Fig. 2 and evaluate the energy involved. We consider the other 5 chains in these hexagons as fixed so that, excluding the flipping chain, hexagon **a** contains $n_a$ Si-chains and $5-n_a$ F-chains and similarly for hexagon **b**. The change in energy is computed by evaluating the interaction of the chains that will be flipped with all the other chains in the two hexagons.

The initial and final energy, $U_i$ and $U_{ii}$, are given by,

$$U_i = n_a M + (5-n_a)F + n_b S + (5-n_a)M, \qquad U_{ii} = n_a S + (5-n_a)M + n_b M + (5-n_a)F \qquad (1)$$

The change in energy $\Delta U = U_i - U_{ii}$ is given by,

$$\Delta U = [S + F - 2M](n_a - n_b). \qquad (2)$$

This indicates clearly that there is only *one* energy, $U = 2M - S - F$, that is relevant in this problem and it is positive in the case where like chains attract most strongly. In what follows all energies are expressed in terms of $U$.

This result occurs because the two chains occur in pairs and an interchange of two chains necessarily changes the numbers of *both* the Si-and F-chains in two neighbouring hexagons. This discussion also highlights the equivalence of the interactions between the Si- and F-chains. Exactly the same ordering occurs if the energy $U$ is dominated by the attraction between the Si-chains, as is expected from chemical arguments, or if the ordering were driven by the attraction between the F-chains or the reduction in the attraction between unlike chains. This means that within the pairwise model the phase shown in Fig. 1(e) should not occur in isolation but should be accompanied by a phase where the F-chains are on an ordered sublattice.

Simple mean field theory [9] predicts that the energy of the ground state, relative to the disordered state, per hexagon is $-\frac{3}{4}U$ and the partially ordered state has energy is $-\frac{1}{4}U$. The entropy of the disordered state is $3k_B\log_e 2$ per hexagon and that of the partially ordered state is $k_B\log_e 2$ per hexagon; the ground state is totally ordered and so has zero entropy.

**IV Monte Carlo Simulations of the model with pairwise interactions.**

The phase diagram for the model that including only the pairwise interactions is determined using Monte Carlo techniques. The simulations were performed using simulations on a periodic 120×120 lattice of hexagons. A terphenyl molecule is selected at random and the energy of the two adjoining hexagons, as shown in Fig. 2 is evaluated, this is $U_i$ then the side chains fixed to this terphenyl molecular are interchanged, this is $U_{ii}$ as given in equation [1]. In the Monte Carlo simulation it is the energy $U_{ii} - U_i$ that determines if this flip is to be accepted using the Metropolis algorithm.

The lattice was broken down into three sublattices, labelled A, B, C. The order parameters for each sublattice are defined as $m_A = \frac{2n_A - 6}{6}$ where $n_a$ is the number of Si-chains in the A$^{th}$ hexagon and similarly for hexagons B and C. The fact that the number of each type of chain is equal and constant requires that $m_A + m_B + m_C \equiv 0$ at all steps in the calculation. 50 separate simulations were run and averaged. Each simulation was performed for 120 million iterations per temperature step and no differences were observed in the phase diagram obtained from the heating and cooling runs.

These simulation studies gave the results shown in Figs. 3(a-d). The phase diagram averaged over 50 runs is shown in Fig. 3(a). Two transitions show up clearly in the specific heat and the sublattice susceptibilities shown in Figs. 3(b) and 3(c); these occur at $k_B T_L = 1.13U$ and $k_B T_H = 1.43U$. There is long range order below $T_L$; the ratio $T_H/T_L$ ~1.27, is considerably smaller than the ratio of 1.75 given by mean field theory [9]. The sublattice susceptibilities show that the higher temperature transition involves all three sublattices but the fluctuations at the lower transition are concentrated on the sublattice that will correspond to $m = 0$ at $T=0$.

Calculations of the entropy were evaluated using $S(T) = \int_0^T \frac{C(T)}{T} dT$ since it was known that $S(0) = 0$. The computed entropy approached $3k_B \log_e 2$ per hexagon in the limit of very high temperatures (not shown) as expected however the computed entropy at the transition that occurs at lower temperatures can be seen from Fig. 3(c) to be $S(1.13U/k_B) \simeq 0.5 k_B$ compared with the mean field result of $k_B \log_3 2 \simeq 0.69 k_B$ [9]. Thus the mean field theory in which only one type of chain was free to fluctuate below the first ordering temperature [9] predicts neither the ratio of the two ordering temperatures nor the value of the entropy at the low temperature transition correctly.

There are two very unusual features in the plot of the average order parameter shown in fig. 3(a). It is clear that the behaviour in the phase between the two phase transitions is highly unusual because the temperature dependence of the order on each sublattice shows concave curvature, indicating that this is not the true order parameter near to the upper critical temperature, and because such large fluctuations occur in the intermediate phase; and it is also unusual, but not unknown, to have two transitions for a system with only one energy parameter [12].

In order to understand this phase diagram we present the result from typical cooling and heating runs in figs. 3(d) and 3(e). The region between the two phase transitions is characterised by very large fluctuations on all three sublattices (these were averaged out in fig. 3(a)). One can see that at each temperature there is likely to be two sublattices that will evolve at low temperatures into $m = \pm 1$ but that actually they are swapping roles mostly with the $m = 0$ sublattice but also with each other. The swapping behaviour occurs both after heating from the well-defined ground state as well as when cooling and ceases below the lower phase transition. This intermediate fluctuating phase is actually reminiscent of the phase found by mean field theory [9] where it was found that the Si sublattice started to order while there was an intermingling between the F and mixed sublattices. In the mean field calculation the symmetry between the F and the Si sublattices was broken artificially but here it is allowed and hence the $m = 0$ sublattice can interchange with *both* the other sublattices. We note that in this fluctuating phase the roles taken by the Si- and F-chains are identical.

The Monte Carlo results for the model where only pairwise interactions are included have indicated that there are three phases, a disordered phase, a strongly fluctuating phase in which the correlations are falling off very slowly, possibly with a power law, and the fully ordered phase.

The fluctuating phase is reminiscent of the partial ordering seen for Ising spins coupled antiferromagnetically on a triangular lattice in the low temperature limit [5,8,13]. There is an intermediate phase where there is no long range order but the model allows the presence of stable vortices up to some critical temperature also described as a locking transition [5,13]. This can occur in the spin model when the sum of the spins on the frustrated sublattice takes the value of one of the non-frustrated sublattices [8] which can force the order on the two sublattices to interchange locally. It may be related to a classical spin liquid, as occurs for Ising spins on a Kagome lattice where there is strong short range, but no long range, order [14].

There is a mapping of our model where spins $S = 3$ are coupled antiferromagnetically on a triangular lattice $S^z = \frac{2n_{Si} - 6}{2}$; $-3 \leq S^z \leq 3$, where the spins must obey Kawasaki dynamics. However the entropy in both the high and low temperature limits to differs between the spin and the chain models: $S_{spin}= k_B\log_e 7$ compared with $S_{chain}= k_B\log_e 8$ as $T \to \infty$ and $S_{spin} > \frac{k_B}{3}(1.0029)\log_e 7$ and $S_{chain}= 0$ as $T \to 0$. [15]. This mapping was not exact because the mapping did not allow for the different multiplicities for different occupations of the chains within a hexagon and also because of the necessity to have Kawasaki dynamics [16].

In Fig. 4 we show how the fluctuating phase can occur. Above the phase transitions all the hexagons have an equal number of Si- and F-chains on average. When ordering commences the hexagons on one sublattice will be more likely to have 4 or 5 Si-chains. Clearly the hexagons on the sublattice that will end up having 3 Si- and F-chains are also fluctuating and if three of them (labelled 1,2,3) get 4 or more F-chains then they can destabilize the site (0) that was initially on the F sublattice as shown in Fig. 4(b). This is the origin of the vortices that lead to the interchange of the sublattices [8]. In the spin model the mixed site takes all values between $\pm S$ and so this instability can exist down to $T=0$ for $S \leq 3$. This cannot occur for our model of the ordering of the liquid crystal because if one site has 6 chains of the same type, $S^z = 3$, its neighbours must have at least one chain of the other type. Hence we see these strong fluctuations only occur for the order parameter, $m$, in the range $0 < m < 0.8$ which corresponds to hexagons having between 3 and 5 Si-chains on average. The second transition corresponds to where these fluctuations are frozen out and true long range order sets in.

In summary the model with pairwise interactions predicts the presence of an intermediate phase in agreement with the mean field theory and the experiment. However because the Si- and F-chains participate on an equal footing it fails to predict the properties of this phase correctly. Furthermore $\beta > 1$ in for $T \sim 1.5$ is unphysical. It is now clear that the pairwise model cannot explain the experimentally observed phase observed, as shown in fig. 1(e), where there is long range order on one sublattice that is rich in Si-chains. Hence in order to make contact with the experiment it is necessary to include effects that break the symmetry between Si- and F-chains.

**V Simulations with three body interactions for Si-chains**

We considered different methods of breaking the symmetry between Si and F-chains which is inherent in the pair interaction model discussed earlier. This would indicate if the results shown in Fig. 3(d) were due to there being two domains one where the Si ordered preferentially and the other where it was the F-chains that ordered. In that case the Si ordered phase could have been stabilised by a small field.

An obvious method is to add an ordering field on one sublattice which will favour Si-chains on that sublattice; a physical justification for this was that it might be a response to ordering occurring on adjacent planes without actually doing a three dimensional simulation . We found that when the ordering field was reduced the system reverted smoothly to results seen when the field was zero. The ordering field smoothed the first phase transition and hence that no comparison could be made with the experiments which measured a clean continuous transition from the disordered phase.

We can also break the symmetry by including three body interactions between the Si-chains. This means that there is an extra stabilising energy if a given hexagon contains three, or more, Si-chains. There are good physical reasons for the presence of this interaction for the X-shaped molecules under study, because the *Si*-chains are more flexible than the F-chains. Thus it is easier for a Si-chain to enter a F-dominated hexagon than for a F-chain to enter a Si-dominated hexagon. Similarly, a pure Si-hexagon is sterically more favoured than a pure

F-hexagon. This is the symmetry breaking used in our simulations. A hexagon containing $n$ Si-chains has an additional attractive energy when $n > 3$ which depends on the number of ways in which three distinct Si-chains can be chosen within that hexagon, $\frac{n(n-1)(n-2)}{6}$. This energy is shown in Table I.

The Monte Carlo calculations are now repeated when the three body term is included for a 120×120 periodic lattice. Each simulation performed 120 million iterations per temperature step and 50 separate simulations were run and averaged, in this case there was no difference between heating and cooling runs. The results, for the order parameters for the three sublattices, specific heat, entropy and sublattice susceptibilities are shown in Fig. 5 where we see that there are now two well separated and very sharp transitions. The plots shown in this figure correspond to the value of the three body attraction between the Si-chains, $u_3$, of $-U/4$. The Si sublattice orders first, as expected, but this occurs at significantly higher temperature than the first transition of the pairwise model as may be seen by compared Figs. 3 and 5 indicating that some of the frustration has been relieved by the inclusion of the three body term. There is now a second phase where the magnetisations on the B and C sublattices satisfy $m_B = m_C = -\frac{m_A}{2}$ before there is a phase transition at a lower temperature to the fully ordered state. This agrees with the partially ordered phase seen in the experiments. One of the interesting features is that the susceptibilities for all three sublattices show the same critical behaviour at the upper transition but sublattice A, which is preferentially occupied by Si-chains, shows a negligible response at the lower transition. The peaks in the susceptibilities and the specific heat are noticeably sharper for the model including three body interactions, indicating that the transitions are defined more clearly.

When the magnitude of the three body interaction was reduced the two transitions approached each other smoothly and for very small values of the three body interaction the plots became identical to those with only pairwise interactions. This means that the fluctuating phase is quite distinct from the partially ordered phase, shown in Fig. 1(e) because we cannot nucleate the partially ordered phase by including a small symmetry breaking term.

**VI Critical exponents**

Qualitative arguments suggest that the critical exponents for the model with three body interactions should be should be those of a three state Potts model because the hexagons with more than 3 Si-chains are singled out. There can only be one such hexagon within each group of three hexagons. So the problem comes down to positioning this on the lattice which is similar to Kr on graphite which is described by a 3 state Potts model which has $\beta = 1/9 \cong 0.11$[17,18].

Experimental data was obtained for the critical experiment $\beta$ from X-ray diffraction experiments. The lattice parameter of the fully and partially ordered phases is $\sqrt{3}$ times that of the disordered phase, as is shown in Fig. 6(a). This gives rise to superlattice peaks as shown in Fig. 6(b). The value of the order parameter exponent, $\beta$ was measured for samples **A2** and **A1** from plots of $\log_{10}(I_{(10)})$ plotted against $\log_{10}(T_c-T)$, as shown in Fig. 7(a). The slope of the fitted lines equals $2\beta$, this gives $\beta_{expt} = 0.18 \pm 0.02$. We were not able to analyse the diffuse scattering to get clear predictions for $\gamma$ and $\nu$. The small value of $\beta$ indicates that the model with three body forces may not have a continuous phase transition. In many other examples including a three body term in the energy may induce a first order transition but this is not universally so in two dimensional models [19].

An approximate value of for $\beta$ was also obtained from the simulation data for the transitions at $T_H$ that is shown in Fig. 7(b); this gave $\beta = 0.12 \pm 0.01$. This value is significantly smaller than that obtained from the experiment but closer to the value suggested by the analogy with the Potts model. It is possible that the larger value of $\beta$ obtained from the experiment occurs because the third dimension is coming into play. More accurate simulation data to obtain other exponents and a more accurate value for $\beta$ appropriate to this interesting model would be very desirable.

**VII Conclusions**

In this paper we used simulations to investigate the phase diagram of a liquid crystal formed from X-shaped molecules which have complex order in planes that are stacked up the third axis. This group of compounds leads to very rich phase diagrams that may be described by interesting two-dimensional statistical models particularly when the two side chains forming part of the X are dissimilar [9, 10, 11]. We have used simulations to investigate the phase diagram where one side chain from each of the six X-shaped molecules occupies a two dimensional hexagon that contains six side chains. The energetics of the ordering are deduced from an understanding of the molecules involved so that the energy of each hexagon depends on the numbers of each type of side chain. The system evolves as the two side chains, that are part of the same X–shaped molecule, are interchanged. The lattice is frustrated as it is not possible for the two types of side chain to segregate completely on the triangular lattice. However, because the numbers of each type of side chain are exactly equal, the system evolves to a unique ground state in which the two of the three sublattices of the triangular lattice are occupied by only one type of side chain and the third contains an equal mixture.

The most interesting feature concerns an intermediate phase that occurs between the disordered phase that exists at high temperatures and the fully ordered phase that exists at low temperatures. Monte Carlo simulations have shown the existence of two possible intermediate phases.

A model using only pairwise interactions between the side chains and in this model the two types of side chain are treated on an equal footing in this model because whatever the choice of the pair interactions between the side chains, there is only a single interaction energy. This model was shown to have a very interesting intermediate phase where substantial amounts of order occurred on one of the three sublattices but the ordering fluctuated between the three sublattices. This is clearly a fascinating dynamic phase that merits further investigation as it appears to be a new form of a classical spin liquid. The experimental data showed clearly that the ordering of the two types of side chain is inequivalent and hence cannot be fitted by any values of the pair interactions.

A refined model was investigated where three body forces were included to break the symmetry between the two types of side chain and this was also studied using Monte Carlo simulations. This led to a different intermediate phase in which there was long range order on only one sublattice. In this case there were two clear phase transitions between the intermediate phase and both the disordered phase and the fully ordered phase. Qualitative arguments were presented to suggest that this phase might be in the universality class for the 3-state Potts model. This intermediate phase was shown to be distinct from the fluctuating phase that occurs where there are only pairwise interactions. This model appeared to fit the experimental results.

We found that this was a novel but also a very rich system. The fluctuating phase that exists over a finite temperature interval appears to be unique. It had some features in common with antiferromagnetically coupled spins on a triangular model and Ising spins on a Kagome lattice except that for the spin models the fluctuations occur all the way down to $T=0$. After the model was modified to include a three body term it appeared to be equivalent to a ferromagnetic three state Potts model. We noted that if each chain only interacted with its nearest neighbours, instead of with all the other five chains in the hexagon, we would get the antiferromagnetic Ising model [9] on a Kagomé lattice or, if three body interactions were included, the three state Potts model also on a Kagomé lattice. Neither of these models order. This means that our physically inspired model, in which we considered the total interaction energy of all the side chains within a hexagon, was actually crucial to getting any ordered phases, as are observed in the experiment.

**Acknowledgements**. We acknowledge financial support from the Leverhulme Foundation, (RPG-2012-804). We thank Prof. C. Tschierske of Martin Luther University, Halle, for supplying the compounds and for many helpful discussions.

**Table 1**

The energies of a hexagon with *n* Si-$_i$ chains and (6 −*n*) F- chains for both pair interactions and including three body interactions and the multiplicities for each choice of *n*.

| Number of Si chains | Number of F chains | Energy including only pair interactions | Additional energy when including 3-body interactions for Si-chains | Multiplicity |
|---|---|---|---|---|
| 6 | 0 | 15S | 20$u_3$ | 1 |
| 5 | 1 | 10S+5M | 10$u_3$ | 6 |
| 4 | 2 | 6S+F+8M | 4$u_3$ | 15 |
| 3 | 3 | 3S+3F + 9M | $u_3$ | 20 |
| 2 | 4 | 6F+S+8M | 0 | 15 |
| 1 | 5 | 10F+5M | 0 | 6 |
| 0 | 6 | 15F | 0 | 1 |

**Figure Captions**

1.(Color on line) (a) The structure of the molecules A1 and A2. (b) Schematic drawing of the liquid crystalline X-shaped molecule. (c) The state where the side chains are as fully ordered as allowed by the frustration. (d) The fully ordered state in terms of hexagons. (e) The experimentally observed partially ordered state, as confirmed by X-ray and neutron diffraction and AFM. (f) The fully disordered phase.

2.(Color on line). The configurations (i) and (ii) of the chain that is flipped between hexagons **a** and **b**.

3.(Color on line). Temperature dependence of the parameters of the pairwise model obtained from Monte Carlo simulations; (a) shows the order parameters, (b) shows the sublattice susceptibilities, (c) shows the specific heat and the entropy; these plots were obtained from the average over 50 heating runs; (d) shows the order parameters in a single cooling run and (e) shows the order parameters in a single heating run.

4.(Color on line). Indicates partial order (by the use of pale colors) in (a) the sites labelled 1, 2, 3 which are mixed (green) fluctuate to blue and this destabilizes the site (0) and the green and blue sublattices start to interchange.

5.(Color on line). The temperature dependence of the parameters of the model that included three body interactions obtained from simulations (a) sublattice magnetisations, (b) heat capacity and the entropy, (c) susceptibility for sublattice A and (d) susceptibility for sublattices B and C.

6.(Color on line). (a) The unit cells for the disordered phase and the partially and fully ordered phases and (b) the X-ray scattering from sample **A1** with $T_c = 80°C$ showing the emergence of the superlattices peaks at low temperature.

7.(Color on line). (a) Intensity of the (10) diffraction peak as a function of temperature, for compounds **A1** and **A2**, together with the fitted curves. The curves are plotted both as $I_{(10)}$ vs $T$ and $\log_{10}(I_{(10)})$ vs $\log_{10}(T_c-T)$ for clarity. The heating rate used is 1°C/min. **A1**, $T_c = 56.3$ °C, $2\beta = 0.36$, and $\beta = 0.18$; **A2**, $T_c = 81.1$ °C, $2\beta = 0.33$ and $\beta = 0.17$; (b) The log-log plots of the order parameter red (dots) obtained from the simulation below $T_H$ fitted to $\beta = 0.12$ using blue(dotted) line.

# Figures

**Figure 1**

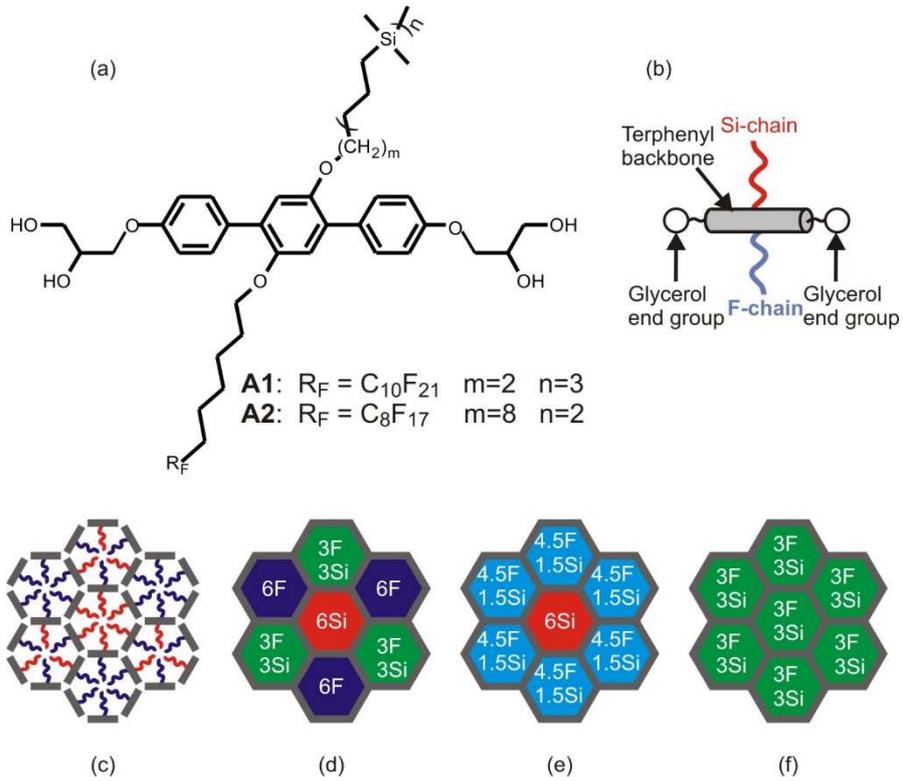

**Figure 2**

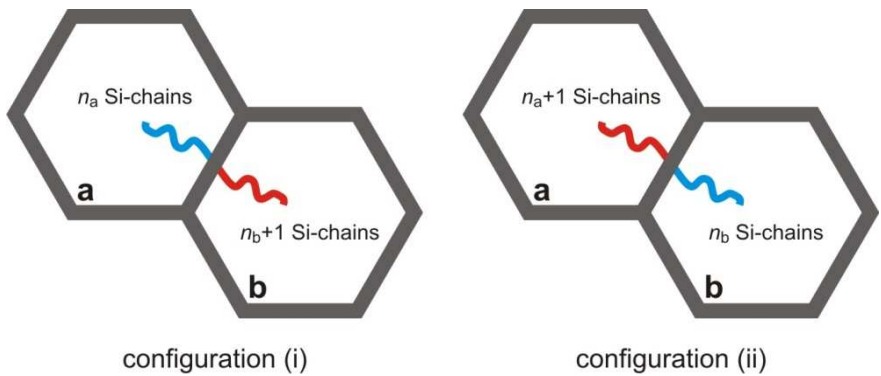

**Figure 3**

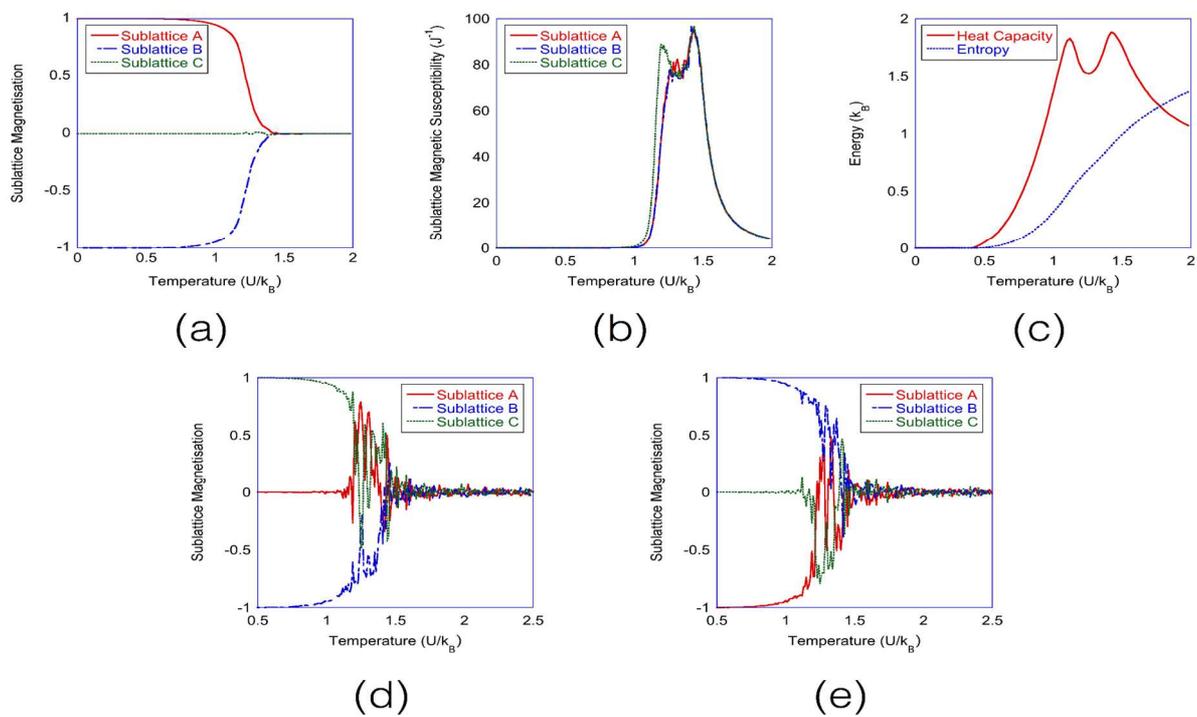

**Figure 4**

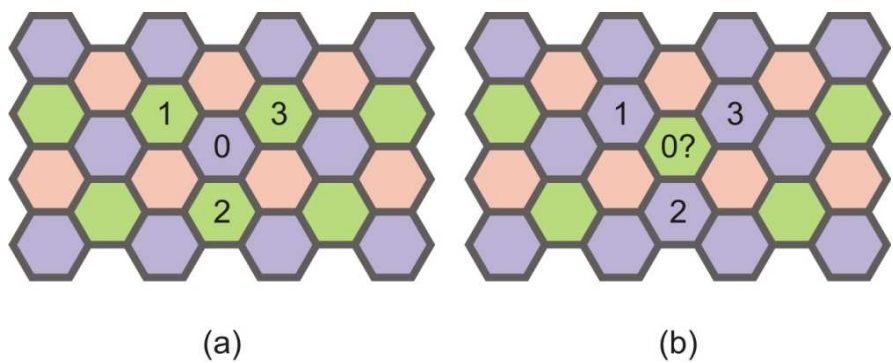

**Figure 5**

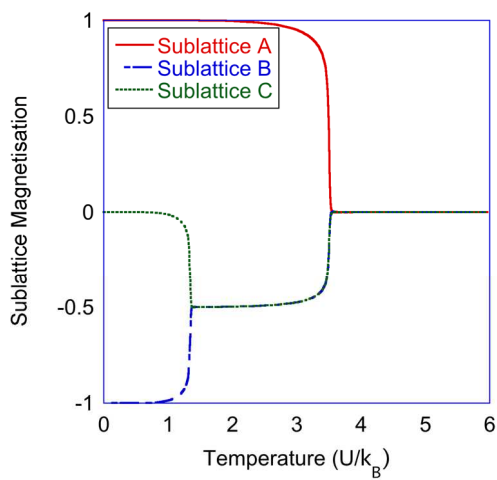
(a)

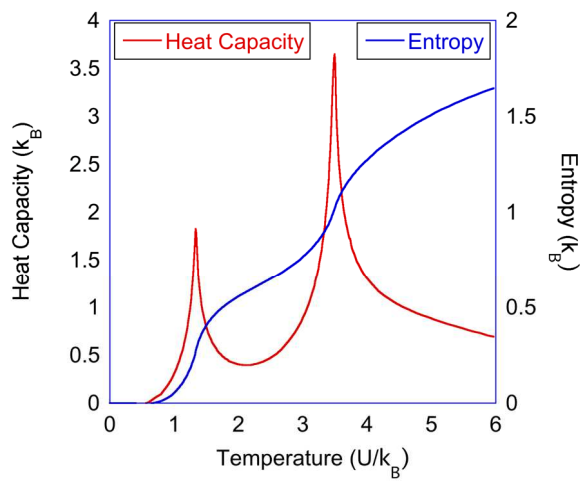
(b)

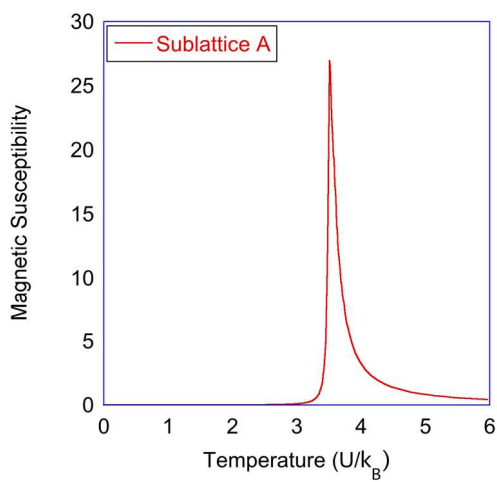
(c)

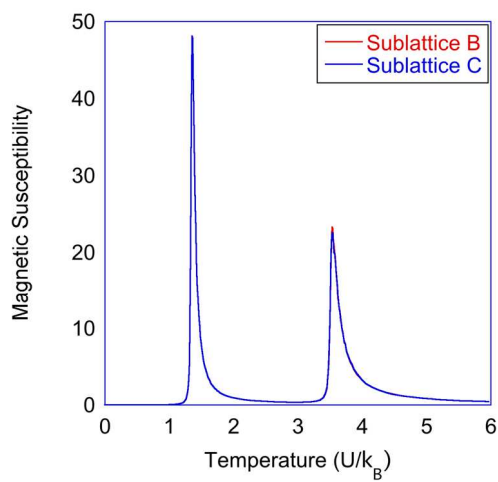
(d)

**Figure 6 (a)**

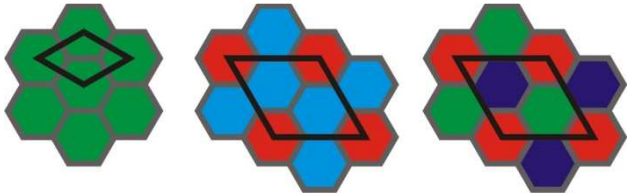

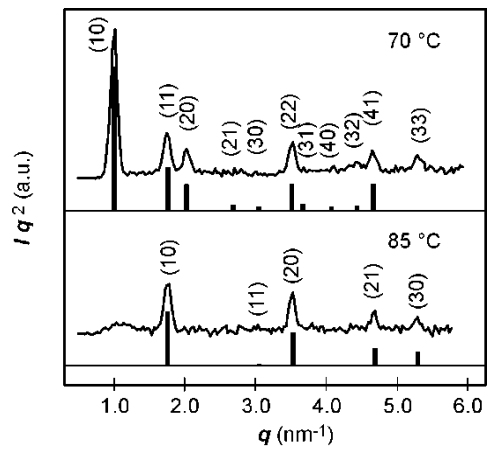

**6(b)**

**Figure 7a**

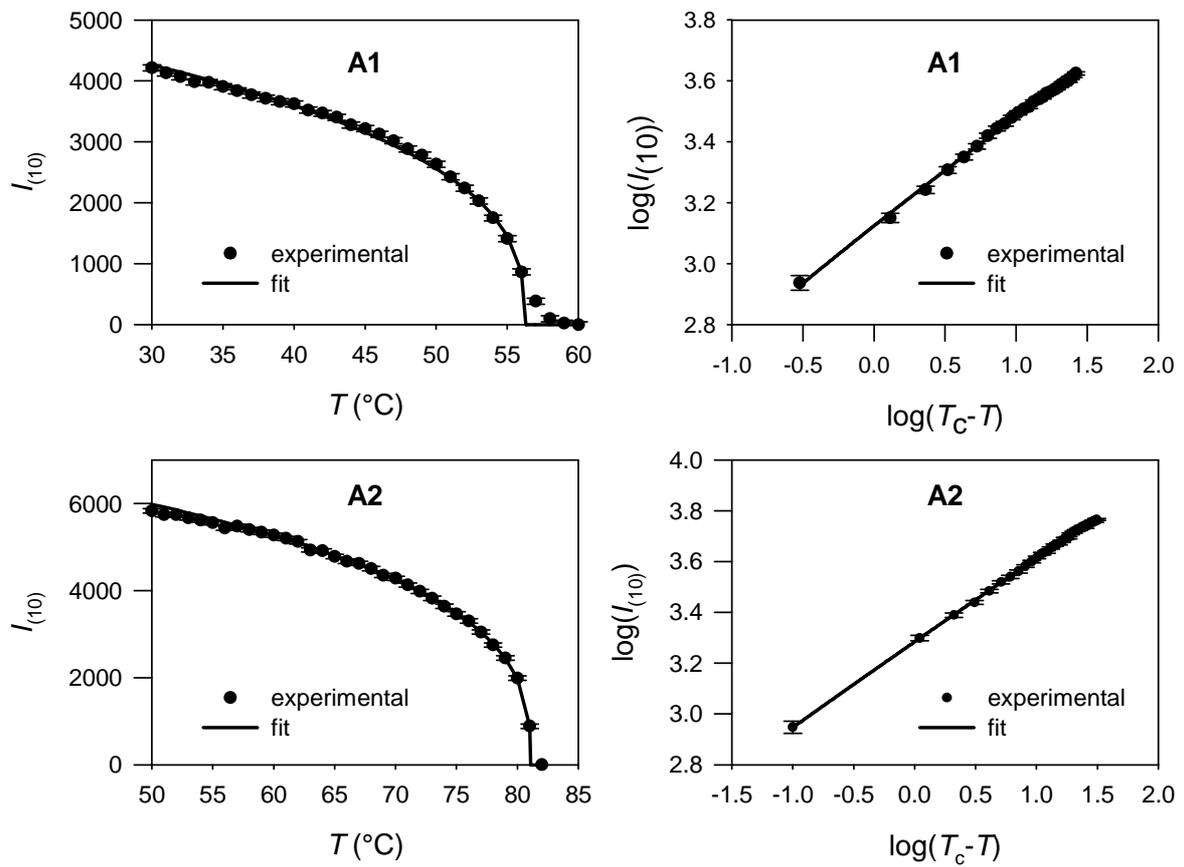

**Figure 7b**

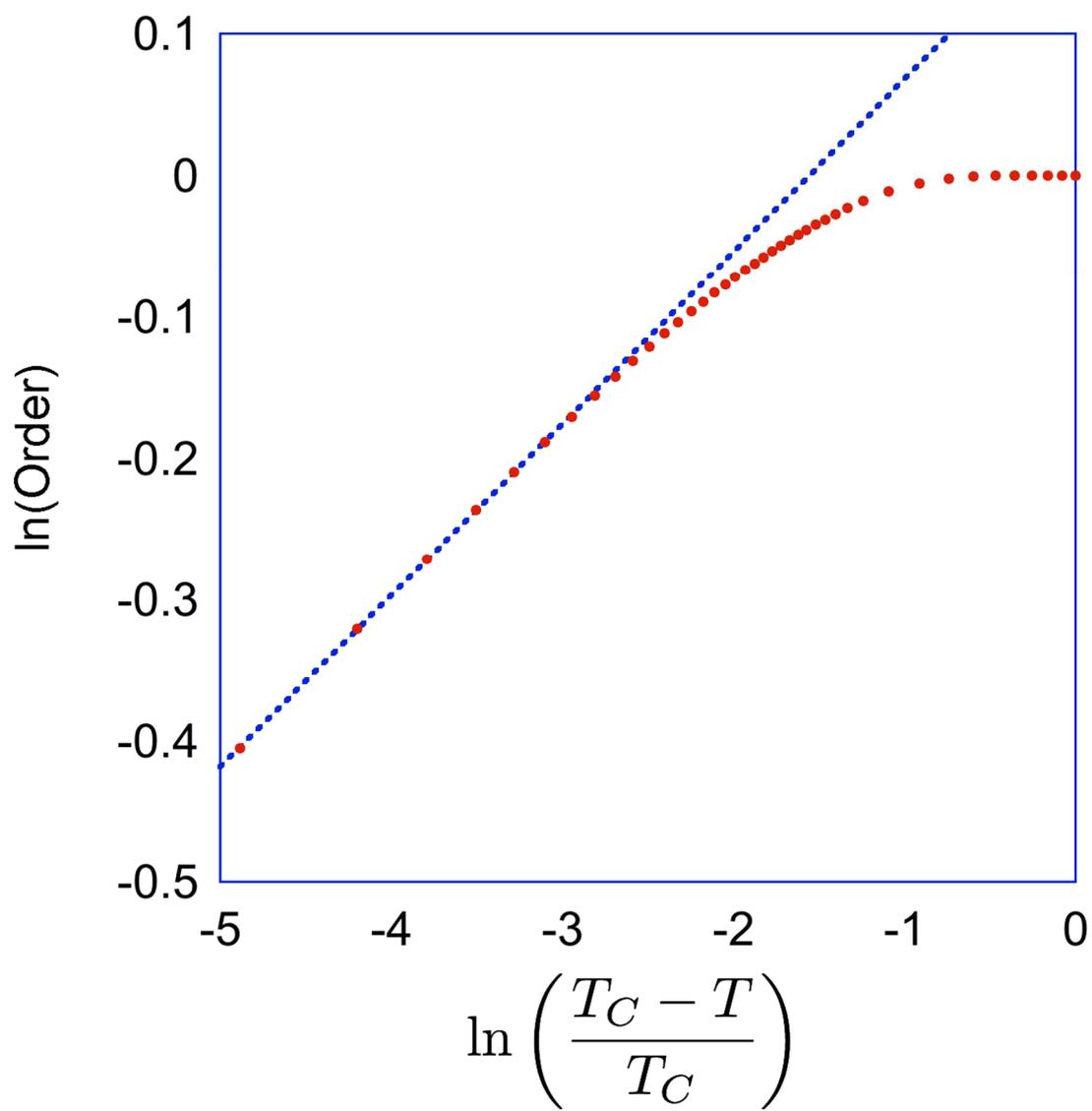